\documentclass{Interspeech2024}

\interspeechcameraready

\title{Fast Context-Biasing for CTC and Transducer ASR models \\
with CTC-based Word Spotter
}

\name{Andrei Andrusenko$^1$, Aleksandr Laptev$^1$, Vladimir Bataev$^{1,2}$, Vitaly Lavrukhin$^1$, Boris Ginsburg$^1$}

\address{
  $^1$NVIDIA\\
  $^2$University of London, UK
}
\email{\{aandrusenko,alaptev,vbataev,vlavrukhin,bginsburg\}@nvidia.com}


\begin{document}

\maketitle


\begin{abstract}
    
Accurate recognition of rare and new words remains a pressing problem for contextualized Automatic Speech Recognition (ASR) systems. Most context-biasing methods involve modification of the ASR model or the beam-search decoding algorithm, complicating model reuse and slowing down inference. This work presents a new approach to fast context-biasing with CTC-based Word Spotter (CTC-WS) for CTC and Transducer (RNN-T) ASR models. The proposed method matches CTC log-probabilities against a compact context graph to detect potential context-biasing candidates. The valid candidates then replace their greedy recognition counterparts in corresponding frame intervals. A Hybrid Transducer-CTC model enables the CTC-WS application for the Transducer model. The results demonstrate a significant acceleration of the context-biasing recognition with a simultaneous improvement in F-score and WER compared to baseline methods. The proposed method is publicly available in the NVIDIA NeMo toolkit\footnote{\scriptsize{\url{https://github.com/NVIDIA/NeMo/blob/main/tutorials/asr/ASR_Context_Biasing.ipynb}}}.
\end{abstract}

\section{Introduction}

ASR models often struggle to recognize words that were absent or had few examples in the training data. Context-biasing methods attempt to solve this problem by assuming that we have a list of words and phrases (context-biasing list) in advance for which we want to improve recognition accuracy.

One of the directions of context-biasing methods is based on the ``deep fusion''. These methods require intervention into the ASR model and its training process. In this case, the context-biasing list is supplied to the encoder or decoder via a cross-attention mechanism as a vector of an entire word \cite{Pundak2018DeepCE, Jain2020ContextualRF, Yang2023PromptASRFC} or a token from context trie \cite{Le2021ContextualizedSE, Harding2023SelectiveBW}. There are also methods based on SpeechLM, when the context-biasing list is fed directly into the prompt for the LLM part of the model \cite{Wang2023SLMBT,Chen2023SALMSL}.

Another direction is methods based on ``shallow fusion''. In this case, the only decoding process is modified. Initially, shallow fusion methods were applied to classic ASR systems by adding new words to the WFST decoding graph \cite{Dixon2012ASW,Hall2015CompositionbasedOR}. Shallow fusion methods are also used for End-to-End ASR. During the beam-search decoding, the hypothesis is re-scored depending on the presence of the current word in the context-biasing list \cite{Zhao2019ShallowFusionEC, Jung2021SpellMN, Galvez2023GPUAcceleratedWB}. It is also possible to combine an end-to-end ASR model with WFST to obtain context-biasing abilities of the classic models \cite{Fox2022ImprovingCR,Zhang2021TinyTA,Andrusenko2022ImprovingOO}. 

Despite the advantages of shallow fusion methods in model reuse, these approaches use beam-search decoding. Processing many alternative hypotheses leads to a significant decoding slowdown even for the Connectionist Temporal Classification (CTC) model~\cite{Graves2006ConnectionistTC}. This problem is considerably worsened in the case of the Transducer (RNN-T) model \cite{Graves2012SequenceTW} since beam-search decoding involves multiple Decoder (Prediction) and Joint networks calculations. Moreover, the context-biasing recognition is limited by the model prediction pool biased toward training data. In the case of rare or new words, the model may not have a hypothesis for the desired word from the context-biasing list whose probability we want to amplify.

\begin{figure}[t]
  \centering
  \includegraphics[scale=0.21]{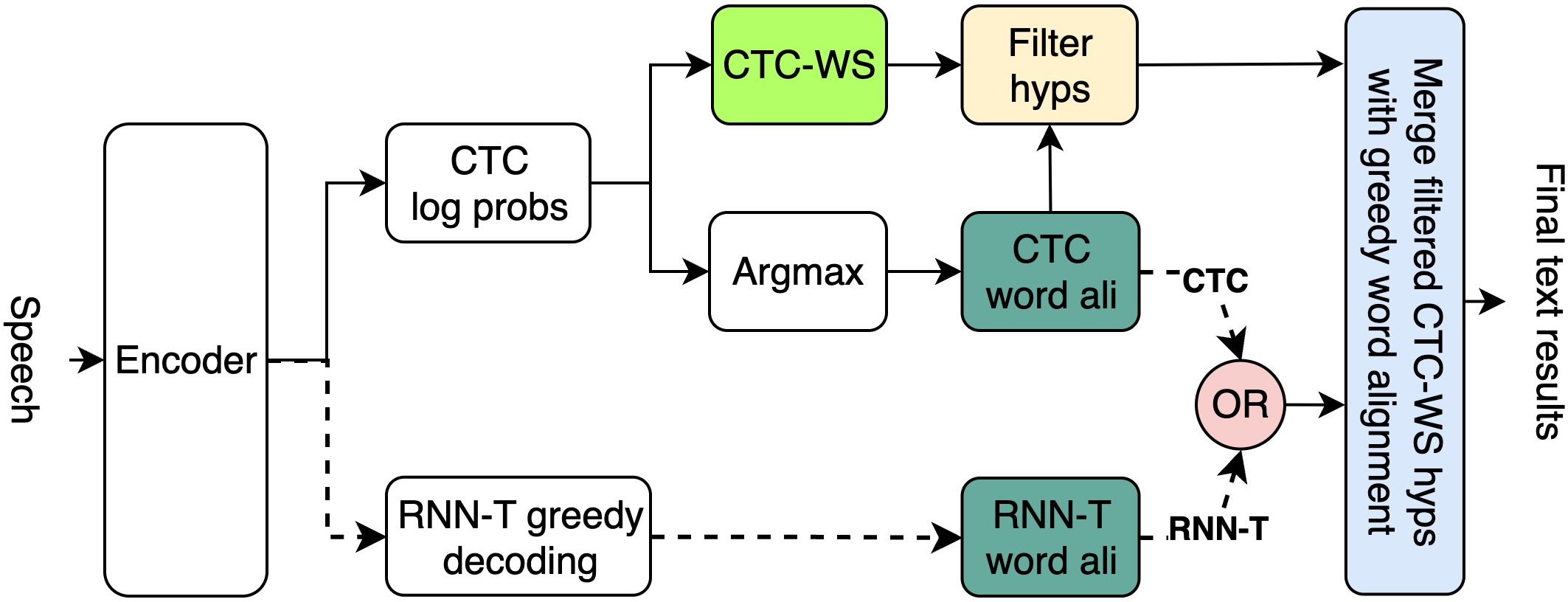}
  \caption{\textit{The proposed context-biasing method. }} 
  \label{fig:ctcws_scheme}
\end{figure}

\begin{figure}[t]
  \centering
  \includegraphics[scale=0.21]{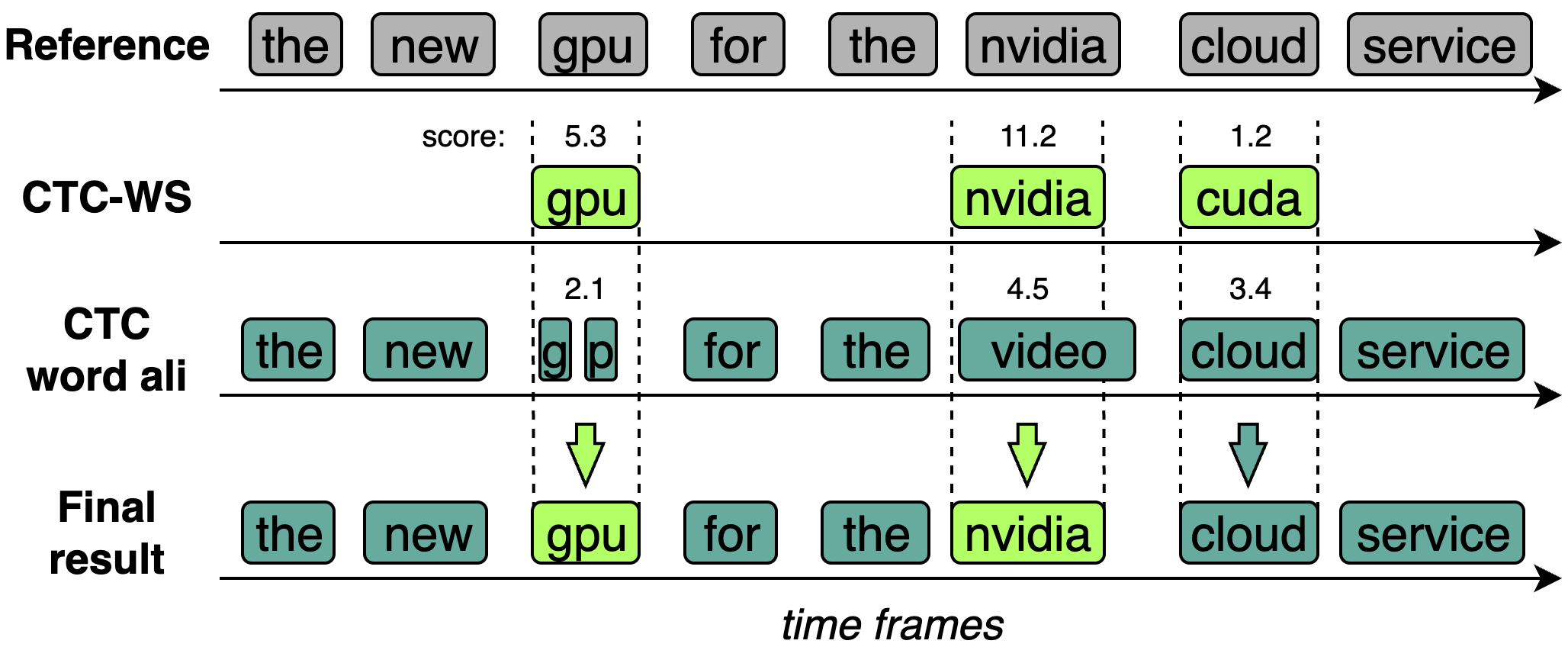}
  \caption{\textit{A context-biasing example for a CTC model.}}
  \label{fig:CTC-WS}
\end{figure}

This work presents a new fast context-biasing method using a CTC-based Word Spotter called CTC-WS (Figure~\ref{fig:ctcws_scheme}).
The method involves decoding CTC log-probabilities with a context graph built for words and phrases from the context-biasing list. The spotted context-biasing candidates (with their scores and time intervals) are compared by scores with words from the greedy CTC decoding results to improve recognition accuracy and reduce false-positive errors of context-biasing (Figure~\ref{fig:CTC-WS}).

We also propose a method of improving the recognition accuracy of abbreviations and complicated words with alternative transcriptions inspired by \cite{Fox2022ImprovingCR}, but obtained automatically without preliminary speech recognition.

A Hybrid Transducer-CTC model \cite{noroozi2024stateful} (a shared encoder trained together with CTC and Transducer output heads) enables the use of the CTC-WS method for the Transducer model. Context-biasing candidates obtained by CTC-WS are also filtered by the scores with greedy CTC predictions and then merged with greedy Transducer results. Compared to baseline shallow fusion methods, the CTC-WS demonstrates better WER and context-biasing word recognition and remarkably speeds up the decoding process for CTC and Transducer models.
The proposed method is publicly available in the NVIDIA NeMo toolkit.

\section{Methods}

\subsection{CTC-based Word Spotter}

\begin{figure}[t]
  \centering
  \includegraphics[scale=0.225]{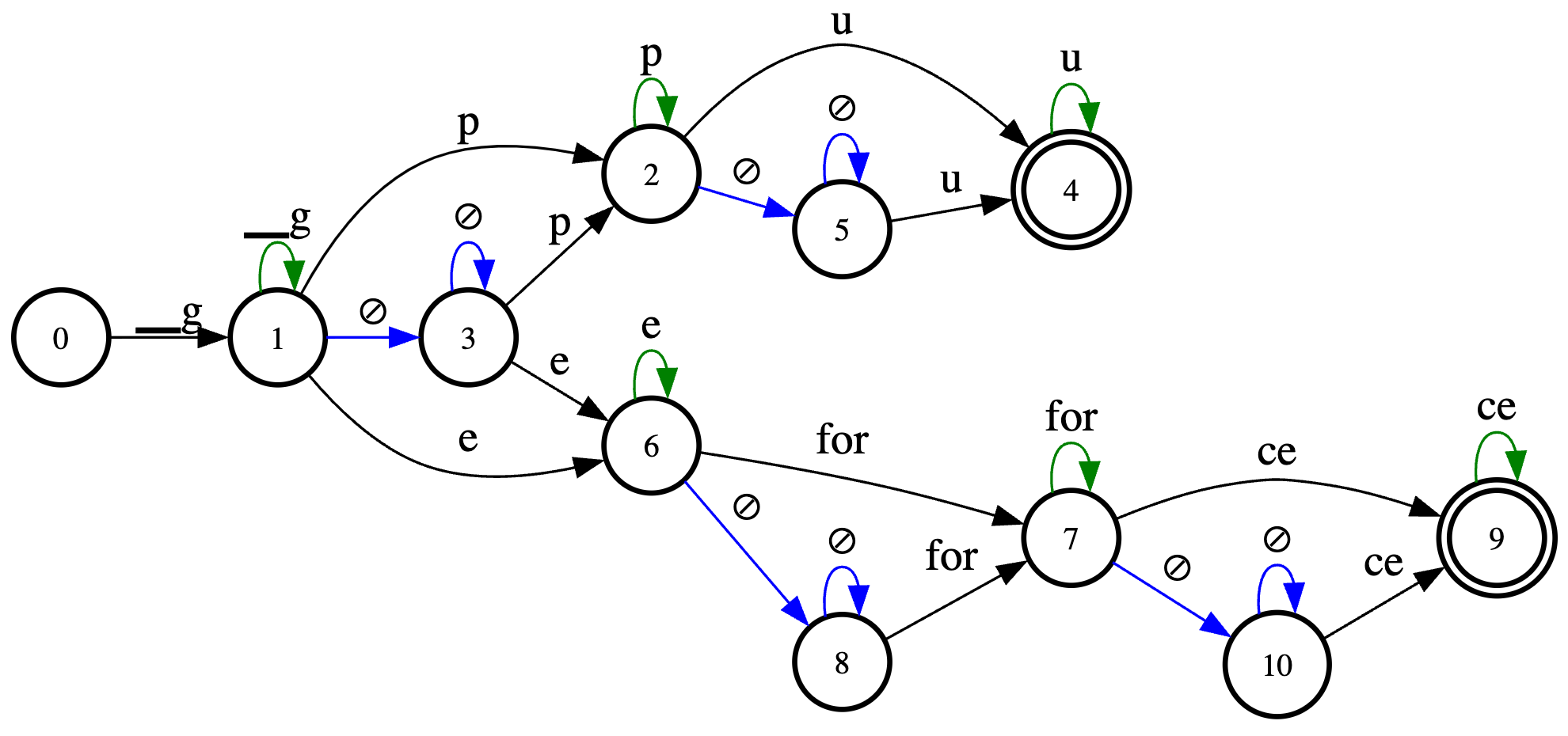}
  \caption{\textit{Context graph -- a composition of a prefix tree with CTC transition topology generated for words ``gpu'' and ``geforce''. Blue and green arcs denote blank ($\varnothing$) transitions and self-loops for non-blank tokens, respectively.}}
  \label{fig:context_graph}
\end{figure}

To solve the context-biasing problem for the CTC model, we propose a new word detection method using CTC-based Word Spotter (CTC-WS).
In the first stage, we build a context graph consisting of a composition of a prefix tree (Trie) with the CTC transition topology for words and phrases from the context-biasing list (Figure \ref{fig:context_graph}). Words are segmented according to the ASR model tokenizer.

Next, we compute log-probabilities (logprobs) from the CTC decoder and perform a decoding process using the context graph according to Algorithm~\ref{alg:ctc-ws}. We add a new empty hypothesis to the root node on every time frame to be able to start recognition of a new word anywhere in the audio.
Hypotheses are additionally rewarded for moving through a non-blank token by adding context-biasing weight $cb_w$ in log-domain, which increases the likelihood of context-biasing words detection.
 
To speed up the decoding process, we use methods for reducing the search space by hypotheses beam and state prunings (line 24 of Algorithm~\ref{alg:ctc-ws}), similar to the classic ASR decoding~\cite{Nolden2011AcousticLF}. As an additional speed-up method, we use a blank skipping technique inspired by~\cite{Zhang2021TinyTA}. If the current hypothesis is empty (at the root state of the context graph) and the probability of the blank output is greater than blank threshold $\beta_{thr}$, we skip this time-frame (line 7). A similar technique can be applied to non-blank tokens with non-blank threshold $\gamma_{thr}$ (line 11).

The word spotter generates candidates of context-biased words with their accumulated scores and frame intervals (start and end positions) in the input audio file. However, detected words may overlap (for example, the same word was recognized with slightly different time shifts). We find all the overlapping intervals and keep only one candidate with the best score on each overlapping (line 27).

Next, we get a word-level alignment of the greedy CTC decoding results (line 28). Words from the alignment are replaced with overlapping context-biasing candidates (line 29). To tackle a false accept problem (the actual word ``cloud'' can be spotted as ``cuda'' if ``cloud'' is not presented in the context graph as per Figure~\ref{fig:CTC-WS} example), we compare the context-biasing candidate score with the score of the overlapping word from the CTC alignment. The word score from the CTC alignment is defined as a sum of logprobs with $ctc_w$ weight for each non-blank token.

\begin{algorithm}[t]
\footnotesize
\caption{CTC-based Word Spotter}\label{alg:ctc-ws}
\begin{algorithmic}[1]
\Require Context graph $CG$, CTC logprobs $L=\{l_0, l_1,..., l_{T-1}\}$, blank threshold $\beta_{thr}$, non-blank threshold $\gamma_{thr}$, context-biasing weight $cb_w$, CTC alignment weight $ctc_{w}$, beam threshold $beam_{thr}$, HYP -- hypotheses class with current $CG$ state, accumulated score, start/end time frames.
\ttfamily
\scriptsize
\State $A = \{\}$ \Comment{list of active hyps}
\State $C = \{\}$ \Comment{list of current hyps}
\State $SH = \{\}$ \Comment{list of spotted hyps}
\For{$t=0$ to $T-1$}
    \State Add HYP$(state=CG.root, start\_frame=t)$ in $A$
    \For{$hyp$ in $A$}
        \If{$hyp$ is empty and $l_t[blank] > \beta_{thr}$}
            \State \textbf{continue}
        \EndIf
        \For{$token$ in $hyp.state.next\_tokens$}
            \If{$hyp$ is empty and $l_t[token] < \gamma_{thr}$}
                \State \textbf{continue}
            \EndIf
            \State $new\_hyp$ $=$ HYP$(state=hyp.state)$
            \State $new\_hyp.start\_frame = hyp.start\_frame$
            \State $new\_hyp.score = hyp.score + l_t[token] + cb_w$
            \If{$new\_hyp.state.is\_end\_of\_word$}
                \State $new\_hyp.end\_frame = t$
                \State Add $new\_hyp$ in $SH$
            \EndIf
            \State Add $new\_hyp$ in $B$
        \EndFor
    \EndFor
    \State $A$ $=$ beam\_and\_state\_prunings($C, beam_{thr}$)
    \State $C = \{\}$
\EndFor
\State $best\_cb\_candidats$ $=$ find\_best\_hyps($SH$)
\State $ctc\_word\_ali$ $=$ get\_ctc\_word\_alignment($L, ctc_{w}$)\\

\Return merge($ctc\_word\_ali, best\_cb\_candidats$)
\sffamily
\end{algorithmic}
\end{algorithm}

\subsection{Hybrid Transducer-CTC model}

The CTC-WS method can also be applied to the Transducer model. To do this, we need an ASR model trained in hybrid mode using a shared encoder with CTC and Transducer decoders, jointly trained with two loss functions (Hybrid Transducer-CTC)~\cite{Huang2022TrainingRW,noroozi2024stateful}.

The only difference with the CTC model is that the final CTC-WS results will have been merged with the results of greedy Transducer decoding. The context-biasing candidates must also be filtered using CTC word-level alignment to avoid a high false accept level. We observed that training such a Hybrid Transducer-CTC model in a joint mode makes the word-level alignment from the CTC and Transducer decoder very close to each other regarding frame intervals, making it possible to apply this method. 

\section{Experimental Setup}

\subsection{ASR model}

As an ASR model, we used the publicly available Hybrid Transducer-CTC\footnote{\scriptsize{\url{https://catalog.ngc.nvidia.com/orgs/nvidia/teams/nemo/models/stt_en_fastconformer_hybrid_large_streaming_multi}}} based on FastConformer encoder architecture \cite{Rekesh2023FastCW} with around 114M parameters. The model was trained on a composite data set of about 20k hours of English speech and a BPE tokenizer \cite{bpe} with 1024 tokens. 

\subsection{Test data}

For a context-biasing benchmark, we collected data from NVIDIA keynote talks. This data is specific to the computer science and engineering domain, which has a large number of unique terms and product names (NVIDIA, GPU, GeForce, Ray Tracing, Omniverse, teraflops, etc.), which is a good fit for the context-biasing task.

All manual transcriptions were normalized using NeMo Text Normalization \cite{Zhang2021NeMoIT} and cleared of remaining non-text characters. All audio files were then segmented from 2 to 35 seconds in duration by the CTC segmentation method \cite{ctcsegmentation}. To address the problem with wrongly aligned segments, we removed segments with a high WER ($>=80$\%) obtained using the baseline ASR model in greedy CTC decoding. The obtained GTC data set was divided into 3-hour dev and 7-hour test sets.

\subsection{Context-biasing list}

While building a context-biasing list, we are interested in high-frequency words and phrases with which the baseline ASR model had recognition problems. We built monogram and bigram-level recognition statistics based on the greedy CTC decoding results to do this. Next, we obtained elements with recognition accuracy $<=50$\%, length $>=3$ characters (short words lead to an excessive number of false accepts and, therefore, should not be included in the context-biasing list), and sorted them according to their frequency. The resulting list was manually processed to remove words like ``okay, gonna, hey, etc.'' which do not fit the context-biasing task. Finally, we obtained a list of 100 unique words and phrases such as \textit{nvidia, geforce, omniverse, tensor core, gpu, cpu,} and others, occurring 739 and 2149 times in dev and test sets, respectively.

We noticed that the baseline ASR model in greedy decoding mode sometimes tends to recognize abbreviations as separate words with single character length (\textit{gpu -- g p u}, \textit{rtx -- r t x}) and compound words as a sequence of words (\textit{hyperscale -- hyper scale}, \textit{tensorrt -- tensor rt}). For the BPE tokenizer, tokens ``\textit{p}'' inside the word ``\textit{gpu}'' and ``\textit{\_p}'' in separate words ``\textit{g p u}'' are different (``\_'' denotes start-of-word in BPE). The ASR model has different decoder outputs for these tokens. To improve the recognition accuracy of such words, we added alternative character-based transcriptions for words $<=4$ character length (presumable abbreviations). For compound words, we use wordninja\footnote{\scriptsize{\url{https://github.com/keredson/wordninja}}}, which can split such words into base words via word statistics from the default wordninja dictionary.

\subsection{Metrics}

To evaluate context-biasing methods, we measure F-score ($2*Precision*Recall/(Precision+Recall)$) for words from the context-biasing list according to the word-level alignment between reference and predicted text. We consider overall WER as well. In addition to accuracy metrics, we measure the time spent on the entire decoding process (with context-biasing), excluding the Encoder pass calculations. All evaluations were carried out on Intel Core i9-10900X and NVIDIA RTX A6000.

\subsection{Baseline context-biasing methods}

Besides greedy CTC and greedy batch Transducer decoding (\textit{batch\_size=128}), we explore publicly available context-biasing methods. The pyctcdecode\footnote{\scriptsize{\url{https://github.com/kensho-technologies/pyctcdecode}}} supports word boosting (\textit{hotwords}) during the CTC beamsearch decoding by shallow fusion approach. We set the default \textit{hotword\_weight} to 10. 

For the baseline beam-search~\cite{Graves2012SequenceTW} and MAES~\cite{Kim2020AcceleratingRT} Transducer decoding methods, we implemented shallow fusion based on the context graph according to the Icefall\footnote{\scriptsize{\url{https://github.com/k2-fsa/icefall/blob/master/icefall/context_graph.py}}} implementation. We used \textit{beam\_size} 5 for all the baseline algorithms because increasing this value further gives a minor accuracy increase while significantly slowing down decoding.

\subsection{CTC-based Word Spotter}

We selected the following CTC-WS parameters that minimize WER without significantly compromising decoding time efficiency obtained for the GTC dev set: context-biasing weight $cb_w = 3.0$, CTC alignment weight $ctc_{w} = 0.5$, blank threshold $\beta_{thr}=\log(0.80)$, non-blank threshold $\gamma_{thr}=\log(0.001)$, and beam threshold $beam_{thr}=7.0$. 

\section{Results}


\begin{table}[ht]
  \centering
  \small
  \caption{\textit{CTC and Transducer decoding results on the GTC test set. CB stands for the presence of context-biasing. Time is overall decoding time without encoder. P is Precision. R is Recall.}}
  \medskip
  \label{tab:ctc_rrnt_results}
  \resizebox{\columnwidth}{!}{%
  \begin{tabular}{l | c | c | c c}
    \toprule
    \textbf{Method} & \textbf{CB} & \textbf{Time, s} & \textbf{F-score (P/R)} & \textbf{WER, \%} \\
    \midrule
    \midrule
    \multicolumn{5}{c}{\textbf{CTC}} \\
    \midrule
    \midrule
    greedy & $\times$ & 3 & 0.32 (0.97/0.19) & 14.02 \\
    \midrule
    \multirow{2}*{pyctcdecode} & $\times$ & 18 & 0.36 (0.98/0.20) & 14.17 \\
     & \checkmark & 179 & 0.79 (0.91/0.69) & 12.06 \\
    \midrule
    CTC-WS & \checkmark & 15 & \textbf{0.87} (0.89/0.85) & \textbf{10.48} \\
    \midrule
    \midrule
    \multicolumn{5}{c}{\textbf{Transducer}} \\
    \midrule
    \midrule
    greedy & $\times$ & 9 & 0.44 (0.98/0.28) & 13.06 \\
    \midrule
    \multirow{2}*{beam-search} & $\times$ & 890 & 0.44 (0.98/0.28) & 12.95 \\
     & \checkmark & 986 & 0.75 (0.90/0.64) & 12.09 \\
    \midrule
    \multirow{2}*{MAES} & $\times$ & 375 & 0.45 (0.98/0.29) & 12.94 \\
     & \checkmark & 453 & 0.80 (0.89/0.73) & 11.39 \\
    \midrule
    CTC-WS & \checkmark & 21 & \textbf{0.87} (0.89/0.85) & \textbf{9.90} \\
    \bottomrule
  \end{tabular}}
\end{table}


\subsection{CTC decoding}

Greedy CTC decoding is fast but has low performance for context-biasing words recognition (Table \ref{tab:ctc_rrnt_results}). This is likely because the model did not have sufficient statistics for such words during training.

Pyctcdecode in the standard beamsearch decoding mode shows almost the same result as greedy decoding. Enabling context-biasing improves F-score and overall WER. However, such context-biasing significantly slows down the decoding process. Increasing the beam size and pruning parameters can slightly improve the accuracy metrics, but the decoding time increases up to 5 times, which is only sometimes justified.

The proposed CTC-WS method demonstrates the most significant improvement in recognition accuracy (F-score and WER) while showing a significant superiority in decoding speed compared to the context-biasing by pyctcdecode due to not running beam-search for decoding.

\subsection{Transducer decoding}

In the case of the Transducer model, greedy decoding works slower than greedy CTC. This is due to multiple calculations by Joint and Prediction networks during decoding. 

Beam-search decoding slows down dramatically, but this allows the use of the context graph in shallow fusion, improving F-score and WER compared to greedy decoding.

MAES reduces decoding time relative to the beam-search by limiting the number of predictions on one audio frame and pruning the hypothesis search space. This approach also works with context graph in shallow fusion. The result obtained is superior to beam-search. This may be due to the improvement in the hypothesis scores according to the rules of prefix search used in MAES.


The proposed CTC-WS approach allows to obtain better F-score (equal to the CTC model results) and WER, while maintaining an advantage in WER over the CTC results. However, the advantage is slightly less than in greedy mode (0.58 instead of 0.96). This may be caused by a better F-score in the greedy mode for the Transducer model. Another factor may be a slight difference in the timestamps of predicted tokens between CTC and Transducer models during context-biasing candidates merging that can affect other words close to the insertion intervals. The overall decoding speed is high since it uses the results of greedy decoding and the fast CTC-based Word Spotter.

We also noticed that despite the relatively close F-score value between the best result of MAES and CTC-WS, the recognition accuracy of some words with non-trivial phonetic transcriptions (for example, nvidia, dlss, kubernetes) is significantly better in the case of CTC-WS. This may be due to the nature of the CTC-WS algorithm. It tries to recognize words based on acoustic conditionally independent predictions from the CTC log-probabilities. MAES, in turn, tries to rescore the already obtained prediction result from the Transducer model. The predictions may not contain the necessary hypotheses because of a limited prediction pool biased toward training data.

\subsection{Ablation study}

\begin{figure}[t]
  \centering
  \includegraphics[scale=0.185]{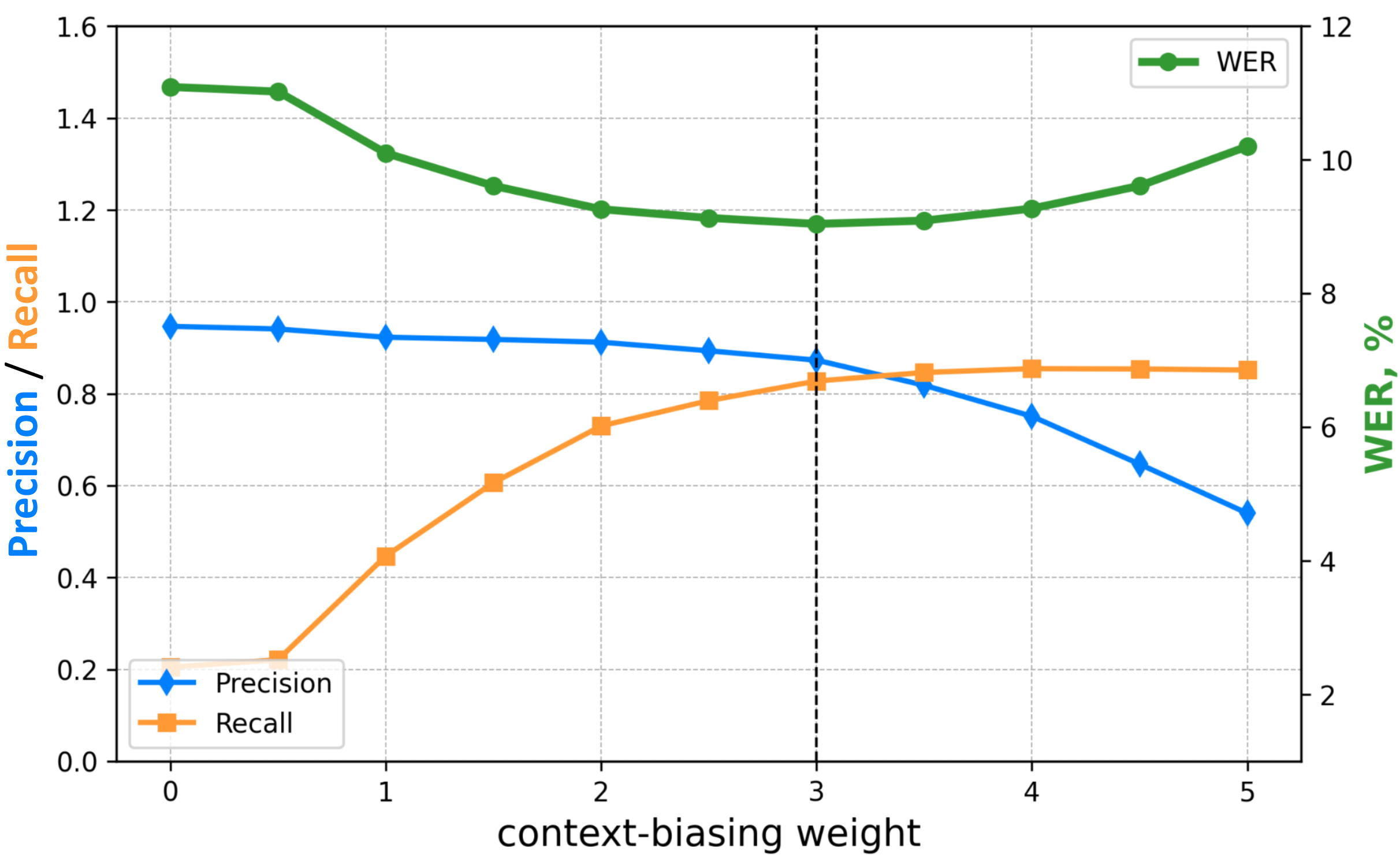}
  \caption{\textit{Precision, Recall, and WER depending on context-biasing weight parameter for the CTC model with CTC-WS and fixed $ctc_{w}=0.5$ for the GTC test set.}}
  \label{fig:prec_rec_wer}
\end{figure}

Figure~\ref{fig:prec_rec_wer} shows recognition quality metrics with respect to context-biasing weight $cb_w$ with fixed CTC alignment weight $ctc_{w} = 0.5$ for the GTC test set. Value $cb_w=3.0$ demonstrates the lowest WER and trade-off \textit{Precision/Recall} ratio.

Automatic alternative transcriptions for short (presumable abbreviations) and compound words significantly improved recognition accuracy (second raw in Table \ref{tab:scalability_test}).
We also explored adding manual transcriptions based on model recognition errors (for example, nvidia -- ``in video'', geforce -- ``g force''). This method allows to get additional improvement in accuracy.
However, the alternative transcriptions can sometimes cause additional false accept errors due to the increased number of candidates during decoding. We also investigated adding alternative transcriptions based on BPE-dropout \cite{provilkovetal2020bpe} as in \cite{Laptev2021DynamicAU}. This method showed similar accuracy performance but led to a more severe slowdown due to the increasing number of branches (alternative transcriptions) in the context graph.

\begin{table}[h!]
  \centering
  \small
  \caption{\textit{Performance of the proposed CTC-WS method for CTC model depending on alternative transcriptions and the size of the cotext-biasing list. Here, ``-a'' and ``+m'' mean no automatic and adding manual alternative transcriptions.}}
  \medskip
  \label{tab:scalability_test}
  \begin{tabular}{l | c | c c}
    \toprule
    \textbf{CB list size} & \textbf{Time, s} & \textbf{F-score (P/R)} & \textbf{WER, \%} \\
    \midrule
    \multicolumn{4}{c}{\textbf{Alternative transcriptions}} \\
    \midrule
    100 & 15 & 0.870 (0.892/0.850) & 10.48 \\
    100 - a & 13 & 0.848 (0.914/0.791) & 11.13 \\
    100 + m & 15 & 0.904 (0.898/0.910) & 10.12 \\
    \midrule
    \multicolumn{4}{c}{\textbf{Scalability}} \\
    \midrule
    100 & 15 & 0.870 (0.892/0.850) & 10.48 \\
    250 & 19 & 0.864 (0.871/0.856) & 10.57 \\
    500 & 21 & 0.852 (0.846/0.858) & 10.76 \\
    750 & 24 & 0.845 (0.834/0.855) & 10.72 \\
    1000 & 26 & 0.845 (0.834/0.856) & 10.73 \\
    \bottomrule
  \end{tabular}
\end{table}

To evaluate a CTC-WS robustness to the size of the context-biasing list, we expanded the baseline list to 1000 elements. We borrowed the missing words (most of them are distractors, i.e., not presented in the GTC data set) from the Earnings benchmark~\cite{Fox2022ImprovingCR}. Based on the results (Table~\ref{tab:scalability_test}, Scalability section), CTC-WS performs relatively stable as the size of the context-biasing list increases. F-score and WER degrade only slightly. However, the decoding time increases as the word spotter processes more words in the context graph. All the ablation studies equally apply to the Transducer model since the whole context-biasing process is performed on CTC predictions.

\section{Conclusion}

We proposed a new fast context-biasing method for CTC and Transducer ASR models with CTC-based Word Spotter (CTC-WS). It only requires CTC log-probabilities to detect words from the context-biasing list. The obtained words are merged with the greedy CTC or Transducer prediction results with almost zero computational cost. We demonstrated a pronounced advantage of the proposed CTC-WS compared to other shallow fusion methods, significantly speeding up the context-biasing process and showing better recognition accuracy of the context-biasing words and overall WER. 
The proposed method is publicly available in the NVIDIA NeMo toolkit.

In future work, we intend to adapt the proposed context-biasing algorithm for streaming recognition mode. Currently, the CTC-WS method requires access to the whole audio file, which is only available in the offline recognition scenario.

\bibliographystyle{IEEEtran}
\bibliography{mybib}

\begin{thebibliography}{10}
\providecommand{\url}[1]{#1}
\csname url@samestyle\endcsname
\providecommand{\newblock}{\relax}
\providecommand{\bibinfo}[2]{#2}
\providecommand{\BIBentrySTDinterwordspacing}{\spaceskip=0pt\relax}
\providecommand{\BIBentryALTinterwordstretchfactor}{4}
\providecommand{\BIBentryALTinterwordspacing}{\spaceskip=\fontdimen2\font plus
\BIBentryALTinterwordstretchfactor\fontdimen3\font minus \fontdimen4\font\relax}
\providecommand{\BIBforeignlanguage}[2]{{%
\expandafter\ifx\csname l@#1\endcsname\relax
\typeout{** WARNING: IEEEtran.bst: No hyphenation pattern has been}%
\typeout{** loaded for the language `#1'. Using the pattern for}%
\typeout{** the default language instead.}%
\else
\language=\csname l@#1\endcsname
\fi
#2}}
\providecommand{\BIBdecl}{\relax}
\BIBdecl

\bibitem{Pundak2018DeepCE}
G.~Pundak, T.~N. Sainath, R.~Prabhavalkar, A.~Kannan, and D.~Zhao, ``Deep context: End-to-end contextual speech recognition,'' \emph{2018 IEEE Spoken Language Technology Workshop (SLT)}.

\bibitem{Jain2020ContextualRF}
M.~Jain, G.~Keren, J.~Mahadeokar, and Y.~Saraf, ``Contextual rnn-t for open domain asr,'' in \emph{Interspeech}, 2020.

\bibitem{Yang2023PromptASRFC}
X.~Yang, W.~Kang, Z.~Yao \emph{et~al.}, ``Promptasr for contextualized asr with controllable style,'' \emph{ICASSP}, 2024.

\bibitem{Le2021ContextualizedSE}
D.~Le, M.~Jain, G.~Keren, S.~Kim \emph{et~al.}, ``Contextualized streaming end-to-end speech recognition with trie-based deep biasing and shallow fusion,'' \emph{Interspeech}, 2021.

\bibitem{Harding2023SelectiveBW}
P.~Harding, S.~Tong, and S.~Wiesler, ``Selective biasing with trie-based contextual adapters for personalised speech recognition using neural transducers,'' \emph{INTERSPEECH 2023}, 2023.

\bibitem{Wang2023SLMBT}
M.~Wang, W.~Han, I.~Shafran \emph{et~al.}, ``Slm: Bridge the thin gap between speech and text foundation models,'' \emph{ASRU}, 2023.

\bibitem{Chen2023SALMSL}
Z.~Chen, H.~Huang, A.~Y. Andrusenko, O.~Hrinchuk, K.~C. Puvvada, J.~Li, S.~Ghosh, J.~Balam, and B.~Ginsburg, ``Salm: Speech-augmented language model with in-context learning for speech recognition and translation,'' \emph{ICASSP}, 2024.

\bibitem{Dixon2012ASW}
P.~R. Dixon, C.~Hori, and H.~Kashioka, ``A specialized wfst approach for class models and dynamic vocabulary,'' in \emph{Interspeech}, 2012.

\bibitem{Hall2015CompositionbasedOR}
K.~B. Hall, E.~Cho, C.~Allauzen, F.~Beaufays, N.~Coccaro, K.~Nakajima, M.~Riley, B.~Roark, D.~Rybach, and L.~Zhang, ``Composition-based on-the-fly rescoring for salient n-gram biasing,'' in \emph{Interspeech}, 2015.

\bibitem{Zhao2019ShallowFusionEC}
D.~Zhao, T.~N. Sainath, D.~Rybach, P.~Rondon, D.~Bhatia, B.~Li, and R.~Pang, ``Shallow-fusion end-to-end contextual biasing,'' in \emph{Interspeech}, 2019.

\bibitem{Jung2021SpellMN}
N.~Jung, G.~min Kim, and J.~S. Chung, ``Spell my name: Keyword boosted speech recognition,'' \emph{ICASSP}, pp. 6642--6646, 2022.

\bibitem{Galvez2023GPUAcceleratedWB}
D.~Galvez and T.~Kaldewey, ``Gpu-accelerated wfst beam search decoder for ctc-based speech recognition,'' \emph{ASRU}, 2023.

\bibitem{Fox2022ImprovingCR}
J.~D. Fox and N.~Delworth, ``Improving contextual recognition of rare words with an alternate spelling prediction model,'' \emph{Interspeech}, vol. abs/2209.01250, 2022.

\bibitem{Zhang2021TinyTA}
Y.~Zhang, S.~Sun, and L.~Ma, ``Tiny transducer: A highly-efficient speech recognition model on edge devices,'' \emph{ICASSP 2021 - 2021 IEEE International Conference on Acoustics, Speech and Signal Processing (ICASSP)}, pp. 6024--6028, 2021.

\bibitem{Andrusenko2022ImprovingOO}
A.~Andrusenko and A.~Romanenko, ``Improving out of vocabulary words recognition accuracy for an end-to-end russian speech recognition system,'' \emph{Scientific and Technical Journal of Information Technologies, Mechanics and Optics}, 2022.

\bibitem{Graves2006ConnectionistTC}
A.~Graves, S.~Fern{\'a}ndez, F.~J. Gomez, and J.~Schmidhuber, ``Connectionist temporal classification: labelling unsegmented sequence data with recurrent neural networks,'' \emph{Proceedings of the 23rd international conference on Machine learning}, 2006.

\bibitem{Graves2012SequenceTW}
A.~Graves, ``Sequence transduction with recurrent neural networks,'' \emph{ArXiv}, vol. abs/1211.3711, 2012.

\bibitem{noroozi2024stateful}
V.~Noroozi, S.~Majumdar, A.~Kumar, J.~Balam, and B.~Ginsburg, ``Stateful conformer with cache-based inference for streaming automatic speech recognition,'' \emph{ArXiv}, 2024.

\bibitem{Nolden2011AcousticLF}
D.~Nolden, R.~Schl{\"u}ter, and H.~Ney, ``Acoustic look-ahead for more efficient decoding in lvcsr,'' in \emph{Interspeech}, 2011.

\bibitem{Huang2022TrainingRW}
G.~Huang, D.~Zhou, S.~Dan, and F.~Huang, ``Training rnn-t with ctc loss in automatic speech recognition,'' \emph{Journal of Applied Mathematics and Computation}, 2022.

\bibitem{Rekesh2023FastCW}
D.~Rekesh, S.~Kriman, S.~Majumdar, V.~Noroozi, H.~Juang, O.~Hrinchuk, A.~Kumar, and B.~Ginsburg, ``Fast conformer with linearly scalable attention for efficient speech recognition,'' \emph{ASRU}, 2023.

\bibitem{bpe}
R.~Sennrich, B.~Haddow, and A.~Birch, ``Neural machine translation of rare words with subword units,'' in \emph{Proceedings of the 54th Annual Meeting of the Association for Computational Linguistics}, 2016.

\bibitem{Zhang2021NeMoIT}
Y.~Zhang, E.~Bakhturina, K.~Gorman, and B.~Ginsburg, ``Nemo inverse text normalization: From development to production,'' in \emph{Interspeech}, 2021.

\bibitem{ctcsegmentation}
L.~K{\"u}rzinger, D.~Winkelbauer, L.~Li, T.~Watzel, and G.~Rigoll, ``Ctc-segmentation of large corpora for german end-to-end speech recognition,'' in \emph{Speech and Computer}, A.~Karpov and R.~Potapova, Eds.\hskip 1em plus 0.5em minus 0.4em\relax Cham: Springer International Publishing, 2020, pp. 267--278.

\bibitem{Kim2020AcceleratingRT}
J.~Kim, Y.~Lee, and E.~Kim, ``Accelerating rnn transducer inference via adaptive expansion search,'' \emph{IEEE Signal Processing Letters}, vol.~27, pp. 2019--2023, 2020.

\bibitem{provilkovetal2020bpe}
I.~Provilkov, D.~Emelianenko, and E.~Voita, ``Bpe-dropout: Simple and effective subword regularization,'' in \emph{Proceedings of the 58th Annual Meeting of the Association for Computational Linguistics}, 2020.

\bibitem{Laptev2021DynamicAU}
A.~Laptev, A.~Andrusenko, I.~Podluzhny, A.~Mitrofanov, I.~Medennikov, and Y.~N. Matveev, ``Dynamic acoustic unit augmentation with bpe-dropout for low-resource end-to-end speech recognition,'' \emph{Sensors (Basel, Switzerland)}, vol.~21, 2021.

\end{thebibliography}

\end{document}